


\font \bigbf=cmbx10 scaled \magstep2

\def\Dirac{{\raise0.09em\hbox{/}}\kern-0.69em D}

\vglue 1.5cm

\centerline {\bigbf On a Noncommutative Extension of Electrodynamics}
\vskip 1.5cm

\centerline {\bf  J. Madore}
\medskip
\centerline {\it Laboratoire de Physique Th\'eorique et Hautes
Energies\footnote{*}{\it Laboratoire associ\'e au CNRS.}}
\centerline {\it Universit\'e de Paris-Sud, B\^at. 211,  \ F-91405 ORSAY}

\vskip 2cm
\noindent
{\bf Abstract:} \ The Maxwell vector potential and the Dirac spinor used to
describe the classical theory of electrodynamics both have components which
are considered to be ordinary smooth functions on space-time. We reformulate
electrodynamics by adding an additional structure to the algebra of these
functions in the form of the algebra $M_n$ of $n \times n$ complex matrices.
This involves a generalization of the notions of geometry to include the
geometry of matrices. Some rather general constraints on the reformulation are
imposed which can be motivated by considering matrix geometry in the limit
of very large $n$. A few of the properties of the resulting models are given
for the values $n=2,3$. One of the more interesting is the existence of
several distinct stable phases or vacua. The fermions can be quark-like in one
and lepton-like in another.

\vfill
\noindent
LPTHE Orsay 92/21

\noindent
May, 1992
\bigskip
\eject

\beginsection{1 Introduction}

In the usual formulation of electrodynamics the Maxwell potential and the
Dirac spinor are constructed with components which lie in the algebra
${\cal C}$ of smooth functions on space-time. We wish to extend the
construction to the algebra ${\cal A} = {\cal C} \otimes M_n$, where $M_n$ is
the algebra of $n \times n$ complex matrices. The Maxwell potential is a
1-form on space-time. We must therefore be able to define differential forms
on the geometric structure defined by ${\cal A}$. This involves generalizing
the notions of geometry to include the geometry of matrices. We give a brief
review of matrix geometry in Section~2. In Section~3 a noncommutative
generalization of the Maxwell-Dirac action is given. There are several
possible generalizations, depending principally on the structure of the
spinors. At the end of Section 3 we shall make some assumptions which reduce
the possibilities to a set of models parametrized uniquely by the integer $n$,
a mass scale $m$ and the analog $g$ of the electric charge. These can be
partially motivated by considering matrix geometry in the limit of very large
$n$, which in a sense which can be made explicit tends to the geometry of the
ordinary 2-sphere. In Section~4 the properties of the models are sketched for
$n = 2,3$.

The models predict the existence of several distinct stable phases. For
$n = 2,3$ they are respectively 2 and 3 in number. In one phase the bosons are
all massless and the fermions appear as quark-like objects. This we call the
hadronic phase. In another phase all bosons are massive except one. This we
call the third phase, for reasons which will become clear in Section 4.
Finally, in the extra phase possible with $n=3$, all but two of the bosons are
massive and the fermions appear as leptons. We call this phase the leptonic
phase. We interpret one massless boson as the photon and in Section~5 we
introduce {\it ad hoc} an extra Higgs field in order to give a large mass to
the other.

\beginsection{2 Matrix Geometry}

We give here a brief review of some of the details of matrix geometry [1,~2].
For an introduction to noncommutative geometry in general we refer to the work
of Connes [3]. An essential element in differential geometry is the notion of
a vector field or derivation. It is an elementary fact of algebra that all
derivations of $M_n$ are interior. A derivation $X$ is therefore necessarily
of the form $X = {\rm ad}\, f$ for some $f$ in $M_n$. The vector space $D_n$
of all derivations of $M_n$ is of dimension $n^2 -1$.

Let $\lambda_a$, for $1 \leq a \leq n^2-1$, be an antihermitian basis of the
Lie algebra of the special unitary group in $n$ dimensions chosen with units
of a mass scale $m$. The product $\lambda_a \lambda_b$ can be written in the
form
$$
\lambda_a \lambda_b = {1\over 2} C^c{}_{ab} \lambda_c +
{1\over 2} D^c{}_{ab} \lambda_c - {1 \over n} m^2 g_{ab}.        \eqno(2.1)
$$
The structure constants $C^c{}_{ab}$ are real and have also units of mass.
The Killing metric is given by $k_{ab} = -C^c{}_{ad}C^d{}_{bc}$. It is related
to $g_{ab}$ by
$$
k_{ab} = 2 n m^2 g_{ab}.                                        \eqno(2.2)
$$
The tensor $k_{ad}C^d{}_{bc}$ is completely antisymmetric. We shall raise and
lower indices with $g_{ab}$. Then $C_{abc}$ is also completely antisymmetric.
We shall normalize the $\lambda_a$ such that $g_{ab}$ is the ordinary
euclidean metric in $n^2 - 1$ dimensions.

The set $\lambda_a$ is a set of generators of $M_n$. It is not a minimal set
but it is convenient because of the fact that the derivations
$$
e_a = {\rm ad}\,\lambda_a                                       \eqno(2.3)
$$
form a basis over the complex numbers for $D_n$. Any element
$X$ of $D_n$ can be written as a linear combination of the $e_a$:
$X = X^a e_a$, where the $X^a$ are complex numbers. The vector space $D_n$ has
a Lie-algebra structure. In particular the derivations $e_a$ satisfy the
commutation relations
$$
[e_a, e_b] =  C^c{}_{ab} \, e_c.                           \eqno(2.4)
$$

We define differential forms on $M_n$ just as one does in the commutative case
[4]. For each matrix $f$ we define the differential of $f$ by the formula
$$
df(e_a) = e_a(f).                                              \eqno(2.5)
$$
This means in particular that
$$
d\lambda^a(e_b) = [\lambda_b, \lambda^a ] = C_{cb}{}^a\lambda^c.
                                                               \eqno(2.6)
$$
We define the set of 1-forms $\Omega^1(M_n)$ to be the set of all elements of
the form $fdg$ or the set of all elements of the form $(dg)f$, with $f$ and
$g$ in $M_n$. The two definitions coincide because of the relation $d(fg) =
f(dg) + (df)g$. The p-forms are defined exactly as in commutative case [1]
with the product given as usual. The set of all differential forms is a
differential algebra.

There is a basis $\theta^a$ of the 1-forms dual to the derivations $e_a$:
$$
\theta^a(e_b) = \delta^a_b.                                     \eqno(2.7)
$$
We have here suppressed the unit matrix which should appear as a factor of
$\delta^a_b$ on the right-hand side. The $\theta^a$ are related to the
$d\lambda^a$ by the equations
$$
d\lambda^a =  C^a{}_{bc}\, \lambda^b \theta^c,                   \eqno(2.8)
$$
and their inverse
$$
\theta^a = m^{-4} \lambda_b \lambda^a d\lambda^b.                \eqno(2.9)
$$
They satisfy the same structure equations as the components of the
Maurer-Cartan form on the special unitary group $SU_n$:
$$
d\theta^a =
-{1 \over 2} C^a{}_{bc} \, \theta^b \theta^c.                    \eqno(2.10)
$$
The product on the right-hand side of this formula is the product in the
algebra of forms. Using the $\theta^a$ the exterior derivative can be written
as $df = e_a \theta^a$. We shall consider the $\theta^a$ as the analog of a
moving frame. They constitute a set of $n^2-1$ elements each of which is an
$(n^2-1) \times n^2$ matrix. Each $\theta^a$ takes in fact $D_n$, of dimension
$n^2-1$, into $M_n$, of dimension $n^2$. The interior product and the Lie
derivative are defined as usual.

{}From the generators $\theta^a$ we can construct a 1-form $\theta$
in $\Omega^1(M)$ which will play an important role in the study of
gauge fields. We set
$$
\theta = - \lambda_a \theta^a.                               \eqno(2.11)
$$
{}From (2.8) we see that it can be written in the forms
$$
\theta = - {1\over nm^2} \lambda_a d\lambda^a =
           {1\over nm^2} d\lambda_a \lambda^a.                \eqno(2.12)
$$
Using $\theta$ we can rewrite (2.9) as
$$
\theta^a = m^{-4} C^a{}_{bc} \lambda^b d\lambda^c
         - n m^{-2} \lambda^a \theta.                        \eqno(2.13)
$$
{}From (2.8) and (2.10) one sees that $\theta$ satisfies the zero-curvature
condition:
$$
d \theta + \theta^2 = 0.                                      \eqno(2.14)
$$
It satisfies with respect to the algebraic exterior derivative the same
condition which the Maurer-Cartan form satisfies with respect to
ordinary exterior derivation on the group $SU_n$.

We shall introduce a metric on $D_n$ by the requirement that the frame $e_a$
be
orthonormal. For $X = X^a e_a$ and $Y = Y^a e_a$ we define
$$
g(X,Y) = g_{ab}X^a Y^b.                                        \eqno(2.15)
$$
To within a rescaling $g(X,Y)$  is the unique metric on $D_n$ with
respect to which all the derivations $e_a$ are Killing derivations.
The first structure equations for the frame $e_a$ and a linear connection
$\omega^a{}_b$ can be now written down:
$$
d\theta^a + \omega^a{}_b \theta^b = \Theta^a.           \eqno(2.16)
$$
We see then that if we require the torsion form $\Theta^a$ to vanish then the
internal structure is like a curved space with a linear connection given
by
$$
\omega^a{}_b = -{1\over 2}  C^a{}_{bc} \,\theta^c.            \eqno(2.17)
$$
The second structure equation defines the curvature form $\Omega^a{}_b$,
which satisfies the Bianchi identities as before.

The complete set of all derivations of $M_n$ is the natural analog of the
space of all smooth vector fields $D(V)$ on a manifold $V$. If $V$
is parallizable then $D(V)$ is a free module over the algebra of smooth
functions with a set of generators $e_{\alpha}$ which is closed under the Lie
bracket and which has the property that if $e_{\alpha} f = 0$ for all
$e_{\alpha}$ then $f$ is a constant function. The matrix algebra $M_n$ has in
general several Lie algebras of derivations $D$ with this property. The
smallest such one, $D_2$, is obtained by considering three matrices
$\lambda_a$ which form the irreducible $n$-dimensional representation of
$SU_2$. These matrices generate the algebra $M_n$. The most general element of
$M_n$ is a polynomial in the $\lambda_a$. The equations
$$
e_a f = 0, \qquad  e_a = {\rm ad} \,\lambda _a, \quad
1 \leq a \leq 3,                                             \eqno(2.18)
$$
imply that $f$ is proportional to the unit element. The set $D_2$
could also be considered as the natural counterpart of a moving frame on a
manifold [5].

With a restricted set of derivations, one can define the exterior
differential exactly as before using equation (2.5). However now the set
of $e_a$ is a basis of $D \subseteq D_n$. The derivations are taken, so to
speak, only along the preferred directions. Equation (2.6) remains valid,
the only change being that the structure constants are those of the algebra
of derivations. A difference lies in the fact that the forms are of course
multilinear maps on the preferred derivations and are not defined on all
elements of $D_n$. The formula (2.7) which defines the dual forms is as
before but the meaning of the expression $\theta^a$ changes. If we choose for
example $D_2$ as the derivations then $\theta^a$ is a $3 \times n^2$ matrix.
It takes the vector space $D_2$ into $M_n$ and it is not defined on the
$n^2 - 4$ remaining generators of $D_n$. Equation (2.8) remains unchanged but
(2.9) and (2.13) will have to be modified.

Consider the case $D = D_2 \subseteq D_n$. Let $e_r$ for $1\leq r \leq n^2-1$
be a basis of $D_n$ and let $e_a$ for $1\leq a \leq 3$ be a basis of $D_2$. We
can choose the $e_a$ to be the first 3 elements of the $e_r$. Then the $SU_2$
structure constants $C^a{}_{bc}$ are the restriction of the $SU_n$ structure
constants $C^r{}_{st}$. Equation (2.1) will be therefore written
$$
\lambda_a \lambda_b = {1\over 2} C^c{}_{ab} \lambda_c +
{1\over 2} D^r{}_{ab} \lambda_r - {1 \over n} m^2 g_{ab}.        \eqno(2.19)
$$
If a basis $J_a$ of $D_2$ satisfies the commutation relations
$[J_a, J_b] = 2i\epsilon_{abc} J^c$ then $J_aJ^a = n^2-1$. On the other hand,
from (2.19) we see that $\lambda_a\lambda^a = -3m^2/n$. If we write then
$$
\lambda_a = -{i\over 2r} J_a,
$$
we find that $r$ is related to $m$ by the equation $12m^2 r^2 = n(n^2-1)$ and
that the $SU_2$ structure constants are given by
$$
C_{abc} = r^{-1} \epsilon_{abc}.
$$

Let $\theta^r$ be the dual basis of $e_r$ and let $\theta^a$ be the dual basis
of $e_a$. We have then 2 possible expressions for $\theta^a$. We have
$$
\theta^{\prime a} = m^{-4} C^a{}_{rs} \lambda^r d\lambda^s
         - n m^{-2} \lambda^a \theta^\prime,                      \eqno(2.20)
$$
with $\theta^\prime$ constructed using $\lambda^r$ and we have
$$
\theta^a = {n\over 3m^2} (r^2 C^a{}_{bc} \lambda^b d\lambda^c
         + \theta \lambda^a),                                      \eqno(2.21)
$$
with $\theta$ constructed using $\lambda^a$. Both definitions satisfy the
equation (2.7). That is, they coincide as $3 \times n^2$ matrices. In equation
(2.17) each of the $\lambda_r$ can be expanded as a polynomial in terms of the
3 elements $\lambda_a$ and using the Leibnitz rule this yields a long
complicated expression for $\theta^{\prime a}$ in terms of the $d\lambda^a$.
In equation (2.21) the expression $d\lambda^a$ is a $3 \times n^2$ matrix but
it has a natural extension to an $(n^2-1) \times n^2$ matrix in which case it
coincides with the definition of $d\lambda^r$ for the first three values of
the index $r$. The two expressions for $\theta^a$ can be compared therefore as
forms on the complete set of derivations. Whereas by construction
$\theta^{\prime a} (e_r) = 0$ for $r \geq 4$, in general the corresponding
equation for $\theta^a$ would not be satisfied.

Using the basis of $D_2$ and its dual we can write the differential of a
matrix $f$ as
$$
df = e_a f \theta^a.                                         \eqno(2.22)
$$
The complete differential is given by $df = e_r f \theta^r$. If $df(e_a) = 0$
then $e_a f = 0$. This means that $f$ is proportional to the unit element and
therefore that $df = 0$. However if $\alpha$ is a general 1-form then the
condition $\alpha(e_a) =0$ does not imply that $\alpha = 0$. For example any
basis element $\theta^r$ for $r \geq 4$ satisfies the equation
$\theta^r (e_a) = 0$. When we consider the restricted set $D_2$ of derivations
we shall choose the algebra of forms to be the differential algebra generated
by the forms (2.21). In this case if $\alpha$ is a 1-form which satisfies the
condition $\alpha (e_a) = 0$ then $\alpha = 0$.

Using the 1-form $\theta$ we can write the differential of a matrix $f$ as
$$
df = - [\theta, f].                                           \eqno(2.23)
$$
If we consider the algebra of all forms as a ${\Bbb Z}_2$-graded algebra then
we can define another $d$ acting on any form $\alpha$ by the formula [3]
$$
d\alpha = - [\eta, \alpha ]                                   \eqno(2.24)
$$
where $\eta$ is some 1-form and the bracket is ${\Bbb Z}_2$-graded. See also
[6] and [7]. If $\eta^2 = -1$ we have $d^2 = 0$. Equation (2.14) becomes
$$
d\eta + \eta^2 = 1.                                            \eqno(2.25)
$$
The definition (2.24) is interesting in that it does not use derivations and
thus can be used when considering the case of more abstract algebras which
have none.

We shall now consider an extension of matrix geometry by considering the
algebra of matrix-valued functions on space-time [2,~8,~9]. Let $x^{\mu}$ be
coordinates of space-time. Then the set $(x^{\mu}, \lambda^a)$ is a set of
generators of the algebra ${\cal A}$ which is the tensor product
$$
{\cal A} = {\cal C} \otimes M_n,                                \eqno(2.26)
$$
of ${\cal C}$ the algebra of smooth real-valued functions on space-time and
$M_n$. The tensor product is over the complex numbers. Let $e_{\alpha} =
e_{\alpha}^{\mu} \partial_{\mu}$ be a moving frame on space-time and $e_a$
with $1 \leq a \leq 3$ a basis of $D_2$. Let $i = (\alpha, a)$. Then
$1 \leq i \leq  7$. We shall refer to the set
$e_i = (e_{\alpha}, e_a)$ as a moving frame on the algebra ${\cal A}$.

For $f \in {\cal A}$ we define $df$ by equation (2.5) but with the index $a$
replaced by $i$. Choose a basis
$\theta ^\alpha = \theta^\alpha_\lambda dx^\lambda$ of the 1-forms on
space-time dual to the $e_{\alpha}$ and introduce
$\theta^i = (\theta^{\alpha}, \theta^a )$ as generators of the 1-forms
$\Omega^1 ({\cal A})$ as a left or right ${\cal A}$-module. Then if we define
$$
\Omega^1_H = \Omega^1({\cal C}) \otimes M_n,  \qquad
\Omega^1_V = {\cal C} \otimes \Omega^1(M_n),
$$
we can write $\Omega^1({\cal A})$ as a direct sum:
$$
\Omega^1({\cal A}) =
\Omega^1_H \oplus \Omega^1_V.                                 \eqno(2.27)
$$
The differential $df$ of a matrix function is given by
$$
df = d_Hf + d_Vf.                                              \eqno(2.28)
$$
We have written it as the sum of two terms, the horizontal and vertical
parts, using notation from Kaluza-Klein theory. The horizontal component is
the usual exterior derivative $d_H f = e_{\alpha} f \theta^{\alpha}$.
The vertical component $d_V$, given by equation (2.22), is purely algebraic
and it is what replaces the derivative in the hidden compactified dimensions.
The algebra $\Omega^*({\cal A})$ of all differential forms is defined as
usual. It is again a differential algebra.

\beginsection{3 The Maxwell-Dirac action}

We shall now write down the analog of the Maxwell-Dirac action in the geometry
defined by the algebra (2.26). We shall identify a connection with an
anti-hermitian element $\omega$ of $\Omega^1({\cal A})$. We saw above
that it can be split as the sum of two parts which we called horizontal and
vertical. We write then
$$
\omega = A + \omega_V,                                             \eqno(3.1)
$$
where $A$ is an element of $\Omega^1_H$ and $\omega_V$ is an element of
$\Omega^1_V$.

In Section 3.1 we introduced a 1-form $\theta$ in
$\Omega^1(M) \subseteq \Omega^1_V$. We shall use this 1-form as a preferred
origin for the elements of $\Omega^1_V$. We write accordingly
$$
\omega_V = \theta + \phi.                                          \eqno(3.2)
$$
The field $\phi$ is the Higgs field.

We have noted previously that $\theta$ resembles a Maurer-Cartan form. Formula
(3.1) with $\phi = 0$ is therefore formally similar to the connection form
on a trivial principal $U_1$-bundle. We have in fact a bundle over a
space which itself resembles a bundle. This double-bundle structure, which is
what gives rise to a quartic Higgs potential as we shall see below, has been
investigated previously, by Manton [10], Harnad {\it et al.} [11], Chapline
and Manton [12], and, more recently, by Kerner {\it et al.} [13] and by
Coquereaux and Jadczyk [14]. The ${\cal A}$-modules which we shall consider
are the natural generalization of the space of sections of a trivial
$U_1$-bundle since $M_n$ has replaced ${\Bbb C}$ in our models. So the $U_n$
gauge symmetry comes not from the number of generators of the module, which we
shall always choose to be equal to 1, but rather from the factor $M_n$ in our
algebra ${\cal A}$.

Let $U_n$ be the unitary elements of the matrix algebra $M_n$ and let
${\cal U}_n$ be the group of unitary elements of ${\cal A}$, considered as the
algebra of functions on space-time with values in $M_n$. We shall choose
${\cal U}_n$ to be the group of local gauge transformations. A gauge
transformation defines a mapping of $\Omega^1({\cal A})$ into itself of the
form
$$
\omega^{\prime} =  g^{-1} \omega g + g^{-1} dg.                   \eqno(3.3)
$$
We define
$$
\eqalign{\theta^{\prime} &= g^{-1} \theta g + g^{-1} d_V g, \cr
A^{\prime} &= g^{-1} A g + g^{-1} d_Hg,}                           \eqno(3.4)
$$
and so $\phi$ transforms under the adjoint action of ${\cal U}_n$:
$$
\phi^{\prime} = g^{-1} \phi g.
$$
It can be readily seen that in fact $\theta$ is invariant under the action of
${\cal U}_n$:
$$
\theta^{\prime} = \theta.                                       \eqno(3.5)
$$
Therefore the transformed potential $\omega^{\prime}$ is again of the form
(3.2).

The fact that $\theta$ is invariant under a gauge transformation means
in particular that it cannot be made to vanish by a choice of gauge. We have
then a connection with vanishing curvature but which is not gauge-equivalent
to zero. If $M_n$ were an algebra of functions over a compact manifold, the
existence of such a 1-form would be due to the non-trivial topology of the
manifold.

We define the curvature 2-form $\Omega$ and the field strength $F$ as usual:
$$
\Omega = d\omega + \omega^2, \quad F = d_H A + A^2.
$$
In terms of components, with
$\phi = \phi_a \theta^a$ and $A = A_{\alpha} \theta^{\alpha}$ and with
$$
\Omega = {1\over 2} \Omega_{ij} \theta^i \wedge \theta^j, \quad
F = {1\over 2} F_{\alpha \beta} \theta^{\alpha} \wedge \theta^{\beta},
                                                                  \eqno(3.6)
$$
we find
$$
\Omega_{\alpha \beta} = F_{\alpha \beta}, \quad
\Omega_{\alpha a} = D_{\alpha} \phi_a ,   \quad
\Omega_{ab} = [\phi_a,\phi_b] - C^c{}_{ab}\,\phi_c.                 \eqno(3.8)
$$

The analog of the Maxwell action is given by
$$
S_B = \int {\cal L}_B,                                            \eqno(3.9)
$$
where
$$
{\cal L}_B = {1\over 4g^2} Tr(F_{\alpha\beta} F^{\alpha\beta})
+ {1\over 2g^2} Tr(D_{\alpha}\phi_a D^{\alpha}\phi^a) - V(\phi).   \eqno(3.10)
$$
The Higgs potential $V(\phi )$ is given by
$$
V(\phi) = -{1\over 4g^2} Tr(\Omega_{ab} \Omega^{ab}).             \eqno(3.11)
$$
It is a quartic polynomial in $\phi$ which is fixed and has no free parameters
apart from the mass scale $m$. The trace is the equivalent of integration on
the matrix factor in the algebra. The constant $g$ is the gauge coupling
constant. We see then that the analog of the Maxwell action describes the
dynamics of a $U_n$ gauge fields unified with a set of Higgs fields which take
their values in the adjoint representation of the gauge group.

The lagrangian (3.10) is the standard lagrangian chosen for all gauge theories
which use the Higgs mechanism. Given a gauge group the theories differ
according to the representation in which the Higgs particles lie and the form
of the Higgs potential. The particular expression to which we have been lead
has been also found by slightly different, group theoretical, considerations
in the context of dimensional reduction by Harnad {\it et al.} [11] and by
Chapline and Manton [12]. What our formalism shows is that the Higgs potential
is itself the action of a gauge potential on a purely algebraic structure. The
$\Omega_{ab}$ are in fact the components of the curvature $\Omega_V$ of the
connection (3.2):
$$
\Omega_V = d\omega_V + \omega_V^2 =
{1\over 2} \Omega_{ab} \theta^a \wedge \theta^b.
$$

The connection determines a covariant derivative on an associated
${\cal A}$-module [3]. See also [7]. Let $H$ be a $M_n$-module. It inherits
therefore a $U_n$-module structure. Define ${\cal H} = {\cal C} \otimes H$.
Then ${\cal H}$ is an ${\cal A}$-module as well as a ${\cal U}_n$-module. The
form of the covariant derivative depends on the module structure of $H$. The
covariant derivative of $\psi \in {\cal H}$ is of the form
$$
D\psi = d\psi + \omega\psi.                                     \eqno(3.12)
$$
The action of $\omega$ on $\psi$ is determined by the action of $U_n$ on $H$.
We have only then to define the vertical derivatives $e_a \psi$ of $\psi$.
Since ${\cal H}$ is a ${\cal A}$-module, for any $f$ in ${\cal A}$ we must
have the relation
$$
e_i (f \psi ) = (e_i f )\psi + f e_i \psi.                       \eqno(3.13)
$$

Suppose that $H$ is a left module. We shall consider only the case
$H = {\Bbb C}^n$. From (3.13) we see that we must set
$$
e_a \psi = \lambda_a \psi.                                      \eqno(3.14)
$$
The action of $U_n$ can only be left multiplication. We find then that
$$
D_a \psi = \phi_a \psi.                                           \eqno(3.15)
$$

Suppose that $H$ is a bimodule. We shall consider only the case $H = M_n$.
{}From (3.13) we see that we must set
$$
e_a \psi = [\lambda_a, \psi].                                    \eqno(3.16)
$$
There are now two possibilities for the action of $U_n$. We can choose $H$ to
be a bimodule with the adjoint action or a left module with left
multiplication. We find then in the first case
$$
D_a \psi = [\phi_a, \psi].                                        \eqno(3.17)
$$
This is invariant under the adjoint action of $U_n$. In the second
case we find
$$
D_a \psi = \phi_a \psi - \psi \lambda_a.                        \eqno(3.18)
$$
This is invariant only under the left action of $U_n$.

With the frame $\theta^i$ which was introduced above the geometry of
the algebra ${\cal A}$ resembles in some aspects ordinary commutative geometry
in dimension 7. As $n\rightarrow \infty$ it resembles more and more ordinary
commutative geometry in dimension 6 and the frame $\theta^i$ becomes a
redundant one in the limit. Let $g_{kl}$ be the Minkowski metric in dimension
7 and $\gamma^k$ the associated Dirac matrices which we shall take to be given
by
$$
\gamma^k = (1\otimes \gamma^\alpha , \, \sigma^a \otimes \gamma^5 ).
$$
The space of spinors must be a left module with respect to the Clifford
algebra. It is therefore a space of functions with values in a vector space
$P$ of the form
$$
P = H \otimes {\Bbb C}^2 \otimes {\Bbb C}^4.
$$

The Dirac operator is a linear first-order operator of the form
$$
\Dirac = \gamma^k D_k,                                            \eqno(3.19)
$$
where $D_k$ is the appropriate covariant derivative, which we must now
define. The space-time components are the usual ones:
$$
D_{\alpha} \psi = e_{\alpha}\psi + A_{\alpha} \psi
+ {1\over 4} \omega_{\alpha}{}^{\beta}{}_{\gamma}
\gamma_{\beta} \gamma^{\gamma} \psi.                             \eqno(3.20)
$$
The $\omega_{\alpha}{}^{\beta}{}_{\gamma}$ are the coefficients of a linear
connection defined over space-time:
$$
\omega_{\alpha}{}^{\beta} =
\omega_{\alpha}{}^{\beta}{}_{\gamma}\theta^{\gamma}.
$$
By analogy we have to add to the covariant derivative given above a term which
reflects the fact that the algebraic structure resembles a curved space with a
linear connection given by (2.17). We make then the replacement [15]
$$
D_a\psi \rightarrow D_a \psi
        - {1\over 8}  C^b{}_{ca} \gamma_b \gamma^c \psi.       \eqno(3.21)
$$

The analog of the Dirac action is given by
$$
S_F = \int {\cal L}_F,                                          \eqno(3.22)
$$
where
$$
{\cal L}_F =  Tr(\bar\psi \Dirac \psi ).                          \eqno(3.23)
$$

We have therefore defined a set of theories which are generalizations of
electrodynamics to the algebra ${\cal A}$. In order to restrict the
generality we shall make three assumptions. First, we shall suppose
there is no explicit mass term in the classical Dirac action. We have already
supposed that the derivations to be used are the algebra $D_2$. Last, we shall
suppose also that $H$ has the module structure which leads to the covariant
derivative defined by (3.18) to which we add the curvature term
as in (3.21). The last two assumptions can be motivated by showing that in the
limit for large $n$ the covariant derivative tends in a sense which can be
made explicit to that used in the Schwinger model [16].

With the restrictions we have a set of classical models which for each
integer $n$ depend only on the coupling constant $g$ and the mass scale $m$
and given by the classical action
$$
S = S_B + S_F                                                      \eqno(3.24)
$$
where $S_B$ is defined by (3.9) and $S_F$ is defined by (3.22).

Different restrictions result in different models [2,~5,~7,~8,~9,~15,~17]. If
one uses the exterior derivative (2.24) one obtains yet different models
[18,~19] but which are similar at least in the bosonic sector. The main
difference lies in the form of the Higgs potential which is in fact closer in
form to that used in the standard electroweak model. The models which we shall
present are not physically viable. One of the reasons is that there are too
many massless bosons. We add in the last Section an extra {\it ad hoc} term in
the lagrangian to eliminate the unwanted ones.

\beginsection {4 Models}

We return now to the action (3.24) in the case $n=2$ [2,~9] and examine the
resulting physical spectrum. The lagrangian (3.10) is a generalization of
the Yang-Mills-Higgs-Kibble lagrangian, with a more elaborate Higgs sector.
The fermions are Dirac fermions which take their values in the space
$M_2 \otimes {\Bbb C}^2$ and the gauge group is $U_2$. There are therefore
four $U_2$ doublets. From (3.11) and the definition (3.8) of $\Omega_{ab}$ we
see that the vacuum configurations are given by the values $\mu_a$ of $\phi_a$
which satisfy the equation
$$
[\mu_a , \mu_b ] - C^c{}_{ab} \mu_c = 0.                        \eqno(4.1)
$$
The number of solutions to this equation is given by the partition function,
the number of ways one can partition the integer $n$ into a set of decreasing
positive integers. Two obvious solutions are $\mu_a = 0$ which corresponds to
the partition $(1, \dots ,1)$ and $\mu_a = \lambda_a$ which corresponds to the
partition $(n)$. If $n=2$ there are no others. Matter can exist then in two
phases. In the symmetric phase all the gauge bosons are massless and three of
them are gluon-like. The fermions are quark-like. In units of
$(1/2{\sqrt 2})m$ there are two doublets of mass 3 and two of mass 5. We call
this phase the hadronic phase. We plan to suppress the $U_1$ component of the
gauge group and reduce it to $SU_2$. There is no photon then and the fermions
are all neutral. In the broken phase, the gauge bosons are all massive if we
suppress the $U_1$ component. The fermions are again neutral but of different
masses. There are now two doublets of mass 5, a doublet of mass 7 and a
doublet of split mass 5 and 7 units. We call this phase the third phase, for
reasons to be made clear below.

In 2+1 dimensions the standard abelian Higgs lagrangian has a 2-phase
structure. The most general renormalizable Higgs potential is a polynomial of
sixth degree which by an appropriate choice of parameters can be put in the
form
$$
V(\phi ) = \lambda \phi^2 (\phi^2 - v^2)^2.
$$
There are therefore two distinct possible stable phases given by $\phi = 0$
and $\phi = v$. In the former the mass of the gauge field is equal to zero. It
corresponds to what we called the hadronic phase, without confinement however
because the gauge group is abelian. In the latter the mass of the gauge field
is not equal to zero. It corresponds to the third phase.

In the case $n=3$ the fermions are Dirac fermions which take their values in
the space $M_3 \otimes {\Bbb C}^2$ and the gauge group is $U_3$. There are
therefore six $U_3$ triplets. Matter can exist now in 3 phases corresponding
to the 3 partitions of 3. In the symmetric phase all the gauge bosons are
massless and eight of them are gluon-like. The fermions are quark-like. In the
units given above they have all masses of the order of one. This is the
hadronic phase. Since we do not wish to interpret the $U_1$ component of the
gauge group as the photon, the fermions are neutral. In the broken phase which
corresponds to the $n=2$ case the gauge bosons are all massive if we suppress
the $U_1$ component. The fermions are then again neutral and again of
different masses. But in the units given above they still have masses of the
order of one. This is the third phase.

The extra phase for $n=3$ we call the leptonic phase. It is given by the
solution to the equation (5.1) of the form
$$
\mu_a =  -{i\over {\sqrt 2}} m \pmatrix{ 0 &     0 \cr
                                         0 &  \sigma_a}.       \eqno(4.2)
$$
As we shall see below, in this phase there are two massless gauge modes. We
must identify one of the corresponding fields with the photon and in the
next Section we shall introduce an extra Higgs field to give a large mass to
the other mode. Define the matrices $\kappa_4$ and $\kappa_5$ by
$$
\kappa_4 = i \pmatrix{1 & 0 & 0 \cr 0 & 0 & 0 \cr 0 & 0 & 0}, \quad
\kappa_5 = {i \over {\sqrt 2}}\pmatrix{0 & 0 &0 \cr 0 & 1 & 0 \cr 0 & 0 & 1}.
$$
We write the gauge potential in the form
$$
A = A^4 \kappa_4 + A^5\kappa_5 + \pmatrix{0 & W^-\cr W^+ & Z}.
                                                                \eqno(4.3)
$$
Here $A^4$ and $A^5$ are ordinary 1-forms, $W^+$ is a 1-form with values in
${\Bbb C}^2$, $W^- = - (W^+)^*$  and $Z$ is a 1-form with values in the Lie
algebra of $SU_2$. In this phase there are therefore 2 charged gauge bosons
and 3 neutral ones. Their masses are given by
$$
m^2_W = {3\over 2} m^2, \qquad m^2_Z = 4 m^2.                    \eqno(4.4)
$$
There are two massless bosons. If we choose the $A^4$ to represent the photon
then the unit of charge is given by $e=g$. All of the 6 triplets of fermions
have again masses of the order of $m$ and these masses are again different
from the corresponding masses in the hadronic and the third phases. Two
triplets have charge 1 and the other 4 are neutral. We write then the spinor
field in the form
$$
\psi =  \pmatrix{  e   &   \mu   &   \tau   \cr
                 \nu_e & \nu_\mu & \nu_\tau \cr
                  l_1  &   l_2   &    l_3   \cr}.                 \eqno(4.5)
$$
Here, $e$, $\mu$ and $\tau$ are charged doublets; each $\nu$ and $l$ is a
neutral doublet.

\beginsection {5 Phenomenolology}

In the models which we are considering the fermions appear naturally with
classical masses of the order of the mass scale $m$, which is also the order
of magnitude of the classical masses of the massive gauge bosons. The coupling
of the Higgs field to the fermions is however not constrained by gauge
invariance and so there is no reason why the corresponding coefficient in
equations (2.6) for example should be set equal to 1. We could have for any
real number $x$:
$$
D_a \psi = x \phi_a \psi - m\,\psi \lambda_a
- {1\over 8r}  \epsilon_{abc} \gamma^b \gamma^c \psi.          \eqno(5.1)
$$
The same argument applies to the curvature terms. Under an arbitrary local
Lorentz transformation $\Lambda$ the spinor $\psi$ transforms to $\psi^\prime
= S^{-1}(\Lambda) \psi$ and the Dirac matrices transform as
$$
\gamma^\alpha \mapsto \gamma^{\prime \alpha} =
S^{-1}(\Lambda) \gamma^\alpha S(\Lambda).
$$
The space-time components $D_\alpha \psi$ of the covariant derivative of
$\psi$  have been constructed so that they transform correctly, or rather
would have had we added the appropriate gravitational term. The same
behaviour must be required of the extra components $D_a \psi$. The
covariant derivative we have used transforms as it should. But in fact each
term transforms correctly [2] and we could have more generally for any real
numbers $x,y$
$$
D_a \psi = x \phi_a \psi - m\,\psi \lambda_a
- {y\over 8r} \epsilon_{abc} \gamma^b \gamma^c \psi.             \eqno(5.2)
$$
There is no way then to fix even the classical masses of the fermions. They
will depend on the mass scale $m$ and the two parameters $x$ and $y$.

We noticed in the last Section that one of the problems of the models which we
considered was an excess of massless gauge bosons. We introduce therefore one
extra {\it ad hoc} Higgs field $\phi^\prime$ coupled in such a way that the
unique massless boson in the hadronic and the third phase as well as the boson
associated with the $U_1$ component of the residual $U_2$ group in the
leptonic phase acquire a sufficiently large mass so as to be neglected. In the
leptonic phase this means that we set $A^5 = 0$. In the hadronic and third
phase this is a trivial operation. We could in fact have simply started from
the group $SU_n$. However it is more pernicious in the leptonic phase. What we
have to do is suppress the $U_1$-component of the unbroken $U_2$ gauge group.
This is not the same as suppressing the $U_1$-component of the original $U_3$
group.

The model with $n=3$ predicts then three phases in each of which the gauge
bosons and the fermions behave differently. There is a photon for example only
in the leptonic phase. The hadronic phase is QCD with the addition of a set of
massive scalars. The leptonic phase is a variant of the electroweak theory
with two charged $W$'s and three $Z$'s. There are also extra fermions
and parity has not been broken. The third phase could only be of astrophysical
interest.

As an application we shall suppose that space-time has divided into
different regions of distinct phase separated by domain walls. We shall
suppose that near us at least the leptonic phase dominates and the hadronic
regions have been so fragmented that they contain only the minimal number of
quarks necessary to form a colorless state. From `outside' in the leptonic
region a hadronic region resembles then a hadron of the bag model [20] with
the domain wall the bag. As in the case of the bag model one would have to
appeal to higher-order quantum effects to compensate the surface energy and
stabilize the configuration.

There is no separate hadronic or leptonic conservation law and only the total
number of fermions is conserved. A lepton differs from a quark exclusively
through its environment. A fermion is defined to be a lepton if it happens to
be in the leptonic phase and it is a quark if it happens to be in the hadronic
phase. A hadron can `decay' then and release 3 fermions which would be seen by
an observer in the leptonic phase as leptons. The life time of the hadron
would depend then on the properties of the domain wall. In a simple
scalar field theory the stability of such domains, supposed as metastable, has
been investigated by Lee and Wick [21]. One characteristic which distinguishes
a quark from a lepton however is its charge since a fermion in the hadronic
phase has necessarily fractional charge when observed from the leptonic phase.

When a photon from the leptonic phase crosses the domain wall and enters the
hadronic phase it splits as the sum of two terms corresponding to the
decomposition
$$
\kappa_4 = {i\over 3} \pmatrix{2 & 0  & 0 \cr
                               0 & -1 & 0 \cr
                               0 &  0 & -1  } + {1\over 3}.
$$
The coefficient of the first term is one of the gluons. We have supposed that
the coefficient of the second term is suppressed. Without going into the
details of the structure of the domain wall it is not possible to give a
quantitative description of how this suppression takes place. The effect would
be however qualitatively the same as the partial suppression of light when it
enters a polarizing medium. Only one of the helicity states propagates; the
other is absorbed. In the present model instead of being absorbed the mode
acquires a very large mass.

The spinor field in the confined phase as seen from the third phase must be
written
$$
\psi =  \pmatrix{u_R & c_R & t_R \cr
                 u_G & c_G & t_G \cr
                 u_B & c_B & t_B \cr},                            \eqno(5.3)
$$
where $u$, $c$ and $t$ stand for respectively the doublets $(u,d)$, $(c,s)$
and $(t,b)$. This form is imposed by the left action of the gauge group
$SU_3$. To calculate the charge of the fermions in the hadronic phase as seen
from the leptonic phase we calculate the eigenvalues of the matrix
$$
Q= {1\over 3}\pmatrix{2 & 0  &  0 \cr
                      0 & -1 &  0 \cr
                      0 & 0  & -1 \cr}.
$$
That is,
$$
Q\psi = {1\over 3}\pmatrix{2u_R & 2c_R & 2t_R \cr
                           -u_G & -c_G & -t_G \cr
                           -u_B & -c_B & -t_B \cr},             \eqno(5.4)
$$
The $SU_3$ gauge invariance appears to have been broken here, and this would
be understandable since the quarks are being observed from the leptonic
region, where the gauge group has been reduced to $SU_2 \times U_1$. The
breaking depends however on an arbitrary choice of form (3.2) of the solution
$\mu_a$ which defines the leptonic phase. That is, although with the choice we
have made the $u_R$ has charge 2/3, we could equally well have made a choice
in which $u_G$ or $u_B$ had charge 2/3. There is a more serious problem
however. Since a hadron must contain a quark of each of the three colors if it
contains only quarks it must appear neutral as seen by an observer in a
leptonic region. Charged hadrons must contain Higgs particles. From the
expression for $Q$ and the fact that the Higgs scalars lie in the adjoint
representation one sees that when they are within a bag and observed from the
outside they have charges $\pm 1$ and 0.

\beginsection {6 Conclusions}

We have presented some of the details of a series of models with a multiple
vacuum structure described by a lagrangian which is a simple
noncommutative extension of the Maxwell-Dirac lagrangian. Each classical
vacuum solution gives rise to a distinct phase with different physical
properties. We have completed the theory in an {\it ad hoc} way by adding
extra Higgs fields to suppress unwanted abelian components of $U_n$ gauge
potentials. We have implicitly supposed that these extra fields are collective
modes and are not part of the classical theory.

The most realistic model contains 3 phases, which we have called
the leptonic, the hadronic and the third phase. The most novel feature of this
model is the fact that quarks and leptons are the same particles but seen in
different phases. Hadrons are regions of space in the hadronic phase separated
by a bag-like domain wall from the dominant leptonic phase, like chunks of ice
floating in water, or rather, like chunks of water floating in ice, since the
leptonic phase is the ordered one. As in the case of the bag model it would
only be possible to obtain a classical solution corresponding to a
multiple-phase configuration by including some of the quantum effects, in the
form for example of a Casimir energy, to stabilize the domain walls between
the two phases. The grand-unification group is $SU_3$ and it unifies quarks
with 3 families of leptons.

The most serious drawbacks of the model are the facts that the hadrons appear
naturally without charge and that no parity-breaking mechanism is present in
the leptonic phase. There are also too many particles in the leptonic phase
and with the wrong masses. Although the three generations appear naturally
they do so in a completely symmetric way, with equal masses.

\beginsection

{\it Acknowledgements}: The author would like to thank J. Stern for
enlightening discussions on the possibility of a lepton-quark duality. He
would also like to thank A. Khare for drawing to his attention the resemblance
with the abelian model.

\beginsection {References}

\item {[1]}
M. Dubois-Violette, R. Kerner, J. Madore, J. Math. Phys. {\bf 31} (1990) 316.

\item {[2]}
M. Dubois-Violette, R. Kerner, J. Madore, Class. Quant. Grav. {\bf 6} (1989)
1709.

\item {[3]}
A. Connes, Publications of the I.H.E.S. {\bf 62} (1986) 257;
`G\'eom\'etrie noncommutative', InterEditions, Paris, 1990.

\item {[4]}
M. Dubois-Violette, C. R. Acad. Sci. Paris {\bf 307} S\'erie I (1988) 403.

\item {[5]}
J. Madore, Int. Jour. of Mod. Phys. A {\bf 6} (1991) 1287.

\item {[6]}
D. Quillen, Topology {\bf 24} (1985) 89.

\item {[7]}
M. Dubois-Violette, R. Kerner, J. Madore, Class. Quant. Grav. {\bf 8} (1991)
1077.

\item {[8]}
M. Dubois-Violette, R. Kerner, J. Madore, Phys. Lett. {\bf B217} (1989) 485.

\item {[9]}
M. Dubois-Violette, R. Kerner, J. Madore, J. Math. Phys. {\bf 31} (1990) 323.

\item {[10]}
N. S. Manton, Nucl. Phys. {\bf B158} (1979) 141.

\item{[11]}
J. Harnad, S. Shnider, J. Tafel, Lett. in Math. Phys. {\bf 4} (1980) 107.

\item {[12]}
G. Chapline, N. S. Manton, Nucl. Phys. {\bf B184} (1980) 391.

\item {[13]}
R. Kerner, L. Nikolova, V. Rizov, Lett. in Math. Phys. {\bf 14} (1987) 333.

\item {[14]}
R. Coquereaux, A. Jadczyk, World Scientific Lecture Notes in Physics {\bf 16}
(1988).

\item {[15]}
J. Madore, Mod. Phys Lett. A {\bf 4} (1989) 2617.

\item {[16]}
H. Grosse, J. Madore, Phys. Lett. {\bf B283} (1992) 218.

\item {[17]}
B.S. Balakrishna, F. G\"ursey, K.C. Wali, Phys. Lett. {\bf B254} (1991) 430;
Phys. Rev. {\bf D44} (1991) 3313.

\item {[18]}
A. Connes, J. Lott, Nucl. Phys Proc. Suppl. {\bf B18} (1989) 29.

\item {[19]}
R. Coquereaux, G. Esposito-Far\`ese, G. Vaillant, Nucl. Phys. {\bf B353}
(1991) 689.

\item {[20]}
A. Chodos, R.L. Jaffe, K. Johnson, C.B. Thorn, V.F. Weisskopf, Phys. Rev.
{\bf D9} (1974) 3471.

\item {[21]}
T. D. Lee, G. C. Wick, Phys. Rev. {\bf D9} (1974) 2291.

\bye